\documentclass{aa}

\usepackage{graphicx,amssymb,natbib}
\usepackage{txfonts}
\usepackage{mathrsfs}

\usepackage{color}

\begin{document}

\def\A{\,\AA\ }
\def\AF{\,\AA}
\def\arcsec{\hbox{$^{\prime\prime}$}}
\newcommand{\etal}{et~al. }
\newcommand{\ca}{\ion{Ca}{ii} 8542~\A }
\newcommand{\fca}{\ion{Ca}{ii} 8542~\AA}
\newcommand{\cmv}{\mbox{{\rm\thinspace cm$^{-3}$}}}
\newcommand{\cmmf}{\mbox{{\rm\thinspace cm$^{-5}$}} }
\newcommand{\D}{\displaystyle}
\newcommand{\eq}[1]{Eq.\,(\ref{#1})}
\newcommand{\fig}[1]{Fig.~\ref{#1}}
\newcommand{\km}{\mbox{{\rm\thinspace km}}}
\newcommand{\kms}{\mbox{{\rm\thinspace km\thinspace s$^{-1}$}}}
\newcommand{\ms}{\mbox{{\rm\thinspace m\thinspace s$^{-1}$}}}
\newcommand{\K}{\mbox{{\thinspace\rm K}}}
\newcommand{\tab}[1]{Table~\ref{#1}}

\newcommand{\ben}{\begin{enumerate}}
\newcommand{\een}{\end{enumerate}}
\newcommand{\bfig}{\begin{figure}}
\newcommand{\efig}{\end{figure}}
\newcommand{\beq}{\begin{equation}}
\newcommand{\eeq}{\end{equation}}
\newcommand{\mbf}{\mathbf}
\newcommand{\vare}{\varepsilon}
\newcommand{\mcal}{\mathcal}
\newcommand{\ep}{\epsilon}
\newcommand{\cs}{\mathcal{S}}
\newcommand{\csv}{\mathcal{S}_{\mathcal{V}}}
\newcommand{\cv}{\mathcal{V}}
\renewcommand{\thefootnote}{\dag}
\def\degr{\hbox{$^{\circ}$}}

\newcommand{\den}{$N_{\rm e}$ }
\newcommand{\fden}{$N_{\rm e}$}
\newcommand{\mic}{$\xi_{\rm t}$ }
\newcommand{\fmic}{$\xi_{\rm t}$}
\newcommand{\Ha}{${\rm H\alpha}$ }
\newcommand{\ha}{${\rm H\alpha}$ }
\newcommand{\fha}{${\rm H\alpha}$}
\newcommand{\haw}{$\pm$0.29\A}
\newcommand{\caw}{$\pm$0.09\A}
\newcommand{\hawt}{$\pm$0.58\A}
\newcommand{\cawt}{$\pm$0.135\A}
\def\degr{\hbox{$^{\circ}$}}

\title{Validation of the magnetic energy vs. helicity scaling in solar magnetic structures}

\author{K.\,Tziotziou\inst{1}
        \and K.\,Moraitis\inst{1} \and M.K.\, Georgoulis\inst{1,2}
        \and V.\,Archontis\inst{3}}

\institute{Research Center for Astronomy and Applied Mathematics (RCAAM), Academy of
Athens, 4 Soranou Efesiou Street, Athens GR-11527, Greece \and Marie Curie Fellow  \and School of Mathematics and Statistics, St. Andrews
University, St. Andrews, KY169SS, UK}

\date{Received  / Accepted }

\titlerunning{Validation of the E-H diagram of magnetic structures}

\authorrunning{Tziotziou \etal}

\abstract
{}{We assess the validity of the free magnetic energy -- relative magnetic helicity diagram for solar magnetic structures.}
{We used two different methods of calculating the free magnetic energy and the relative magnetic helicity budgets: a classical, volume-calculation nonlinear force-free (NLFF) method applied to finite coronal magnetic structures and a surface-calculation NLFF derivation that relies on a single photospheric or chromospheric vector magnetogram. Both methods were applied to two different data sets, namely synthetic active-region cases obtained by three-dimensional magneto-hydrodynamic (MHD) simulations and observed
active-region cases, which include both eruptive and noneruptive magnetic structures.}
{The derived energy--helicity diagram shows a consistent monotonic scaling between relative helicity and free energy with a scaling index 0.84$\pm$0.05 for both data sets and calculation methods.
It also confirms the segregation between noneruptive and eruptive active regions and the existence of thresholds in both free energy and relative helicity for active regions to enter eruptive territory.}
{We consider the previously reported energy-helicity diagram of solar magnetic structures as adequately validated and envision a
significant role of the uncovered scaling in future studies of solar magnetism.}


\maketitle

\section{Introduction}
     \label{S-Introduction}

Both current-carrying (i.e. ``free'') magnetic energy and magnetic helicity are considered to play an important role in active region (AR) evolution and dynamics \citep{lab07,gtr12,tgr12,tgl13}. In ARs, considerable magnetic flux emergence \citep[$\sim$ 10$^{22}$ Mx,][]{schr:harv} tends to build up strong opposite-polarity regions, sometimes separated by highly sheared polarity inversion lines. Such regions deviate strongly from the ``ground'' current-free (potential) energy configuration, and free magnetic energy quantifies the excess energy on top of this ``ground'' energy state. It builds up through the continuous flux emergence on the solar surface and, to a lesser extent, by other
processes, such as coronal interactions \citep{gals00} or photospheric twisting \citep[e.g.,][]{pari09}. Magnetic free energy is released in the course of solar eruptions (flares and coronal mass ejections [CMEs]) and of smaller-scale dissipative events (subflares, jets, etc.). Magnetic helicity, on the other hand, quantifies the distortion (twist, writhe) and linkage of the magnetic field lines compared to their potential-energy state \citep[e.g.,][]{moff92,berg99}. It mainly emerges from the solar interior via helical magnetic flux tubes or is generated
by solar differential rotation, photospheric turbulent shuffling, or peculiar motions in ARs. Contrary to free energy, magnetic helicity cannot be efficiently removed by magnetic reconnection in high magnetic Reynolds-number plasmas \citep{berg84}, and if not transferred during reconnection events to larger scales via existing magnetic connections, it has to be bodily expelled from the AR in the form of CMEs \citep{low94, devo:00}.

Derivation of the instantaneous free magnetic energy and relative (with respect to the potential field in a closed volume) magnetic helicity budgets in an observed AR is mainly based on temporal integration of an energy/helicity injection rate \citep{berfie84} or on evaluation of classical analytical formulas \citep{finn85, berg99} with the required three-dimensional magnetic field derived from extrapolations of an observed lower photospheric boundary. Both methods involve significant uncertainties and ambiguities,
stemming respectively from the dependence of energy/helicity injection rates on the determination of the photospheric
velocity field \citep[e.g.,][]{wels07} and the model-dependent nonlinear force-free
(NLFF) field extrapolations \citep[e.g.,][ and references therein]{schr06,metc08}. However, for a fully known three-dimensional
magnetic field configuration, self-consistently derived with magneto-hydrodynamic (MHD) numerical simulations, analytical formulas should provide a correct estimate of free-energy/helicity budgets. In this case, possible uncertainties mainly rise from the assumptions adopted in the calculations of the potential field and the calculation of respective vector potentials. Typically, the potential field in localized solar structures (e.g., ARs) is calculated in the semi-infinite volume above a planar lower boundary \citep[e.g., ][]{schm,devo:00}.
Such calculations were recently revised by \cite{mor14} to infer the potential field in a finite volume filled by a nonpotential
field, while \cite{val12a,val12b} revised the calculation of the vector potential in finite, bounded volumes.
These developments enabled the precise volume calculation of magnetic energy and relative magnetic helicity budgets in three-dimensional MHD models of solar magnetic structures via the classical formulation.

In reality, however, since the three-dimensional coronal magnetic structure of ARs is unknown, \citet{gtr12} proposed a new NLFF method that calculates the instantaneous energy and helicity budgets from a single (photospheric or chromospheric) vector magnetogram and used it to study the evolution of these quantities in solar ARs and their role in AR dynamics \citep{tgr12,tgl13}. This method relies on the magnetic connectivity inferred by applying physical arguments to observed photospheric or chromospheric magnetic structures and has been recently validated and benchmarked with three-dimensional coronal structures from MHD models and extrapolations of observed AR magnetograms \citep{mor14}. A major finding of the method, resulting from its application to tens of ARs, is the so-called energy-helicity (EH) diagram of solar ARs \citep{tgr12}. This diagram shows a nearly monotonic dependence between the two quantities and the existence of thresholds of $4 \times 10^{31}$ erg and $2 \times 10^{42}$ Mx$^2$ for free energy and relative helicity, respectively, for an AR to enter eruptive territory. This EH diagram has also been verified with a timeseries of magnetograms of a single AR  \citep[NOAA AR 11158,][]{tgl13} and has also been found to hold for quiet-Sun regions \citep{tz14}. However, the EH diagram has not yet been validated with a
different method for deriving the free magnetic energy and relative magnetic helicity budgets.

In this study we are using both the classical formulas and the recently proposed NLFF method of \cite{gtr12} for both synthetic MHD-derived ARs and extrapolations of observed ARs to further assess the validity of the EH diagram, hence the previously reported monotonic scaling between free magnetic energy and relative magnetic helicity. Section~\ref{S-meth} briefly describes the methodology used, with data and results respectively discussed in Sections~\ref{S-data} and \ref{S-enhel}, while Section~\ref{S-discuss} summarizes and discusses our findings.

\section{Methodology}
\label{S-meth}

We employ two methods for calculating the instantaneous free magnetic energy and relative magnetic helicity budgets of bounded, three-dimensional magnetic structures. The first method is the connectivity-based NLFF method introduced by \cite{gtr12}, which requires a single photospheric or chromospheric vector magnetogram for deriving a unique magnetic-connectivity matrix that relies on a magnetic flux-partition solution for this magnetogram. Such a matrix is derived by means of a simulated annealing method \citep{geo:rus} and contains the flux committed to connections between positive- and negative-polarity flux partitions. Nonzero flux elements of this connectivity matrix define a collection of $N$ magnetic connections, treated as slender force-free flux tubes with known footpoints, flux contents, and variable force-free parameters $\alpha$. The free magnetic energy, which represents a lower limit \cite[see][]{gtr12}, and the respective relative magnetic helicity $H$ are given by
\begin{eqnarray}
E_\mathrm{c}  & = &  A \lambda^2 \sum _{l=1}^N \alpha _l^2 \Phi_l^{2 \delta} +
      \frac{1}{8 \pi} \sum _{l=1}^N \sum _{m=1, l \ne m}^N
           \alpha _l \mathcal{L}_{lm}^{\rm arch} \Phi_l \Phi_m   \label{Ec_fin} \\
H  & = & 8 \pi \lambda^2 A
\sum _{l=1}^N \alpha _l \Phi_l ^{2 \delta} +
      \sum _{l=1}^N \sum _{m=1,l \ne m}^N \mathcal{L}_{lm}^{\rm arch} \Phi_l
      \Phi_m\;\;,
\label{Hm_fin}
\end{eqnarray}
where $\lambda$ is the pixel size of the magnetogram, $A$ and $\delta$ are known fitting constants,  $\Phi_l$ and $\alpha_l$ are the respective unsigned flux and force-free parameters of flux tube $l$, and $\mathcal{L}_{lm}^{\rm arch}$ is the mutual-helicity factor describing the interaction of two arch-like flux tubes \citep{dem06,gtr12}. Details of the method, parameters, and respective uncertainties are described in \cite{gtr12}.

The second method uses the analytical expressions for free magnetic energy and relative magnetic helicity for a three-dimensional magnetic field
$\mathbf{B}$ occupying a finite, bounded volume $\mathcal{V}$, relative to the current-free (potential) magnetic field $\mathbf{B}_\mathrm{p}$, given by
\begin{eqnarray}
E_\mathrm{c} & = & \frac{1}{8\pi}\int_\mathcal{V} \mathrm{d}V\,(\mathbf{B}^2-\mathbf{B}_\mathrm{p}^2)
      \,\,\, = \,\,\, \frac{1}{8\pi}\int_\mathcal{V}\mathrm{d}V\,(\mathbf{B}-\mathbf{B}_\mathrm{p})^2
 \label{ecdef} \\
H & = & \int_\mathcal{V} \mathrm{d}V\,(\mathbf{A}+\mathbf{A}_\mathrm{p})\cdot(\mathbf{B}-\mathbf{B}_\mathrm{p})
\label{hmdef}
\end{eqnarray}
where $\mathbf{A}$, $\mathbf{A}_\mathrm{p}$ are the generating vector potentials of the fields $\mathbf{B}$ and $\mathbf{B}_\mathrm{p}$, respectively.
From the two equivalent expressions for the free energy, the first is used, however if it yields a negative value, then the mean of the two expressions is taken. The potential field $\mathbf{B}_\mathrm{p}$ is equal to $-\nabla \varphi$, where $\varphi$ is a scalar potential, and it is derived by solving Laplace's equation $\nabla^2\varphi=0$ in $\mathcal{V}$ under the Neumann boundary conditions
\begin{equation}
\left.\frac{\partial\varphi}{\partial\hat{n}}\right|_{\partial \mathcal{V}}=-\left.\hat{n}\cdot\mathbf{B}\right|_{\partial \mathcal{V}}\;.
\end{equation}
Under such boundary conditions, $\mathbf{B}$ and $\mathbf{B}_\mathrm{p}$ have the same normal components on all boundaries $\partial \mathcal{V}$ of the finite volume $\mathcal{V}$,  contrary to the semi-infinite solution of \cite{schm}, where this holds only at the lower boundary.
Derivation of the corresponding vector potentials $\mathbf{A}$ and $\mathbf{A}_\mathrm{p}$ is achieved with the method proposed by \cite{val12a, val12b} as implemented by \cite{mor14}. Details of the aforementioned analytical method, the chosen integration methods, and a derivation of uncertainties are provided in \cite{mor14}. While the three-dimensional MHD simulations used here have a given normal-field solution on all boundaries of the simulation
volume, this solution is not a priori known for the extrapolated ARs. In these cases the lateral- and top-boundary normal-field
solutions are provided by the NLFF field extrapolation used (see Section~\ref{S-data} below).

\begin{figure*}
\centering
\includegraphics[width=18cm]{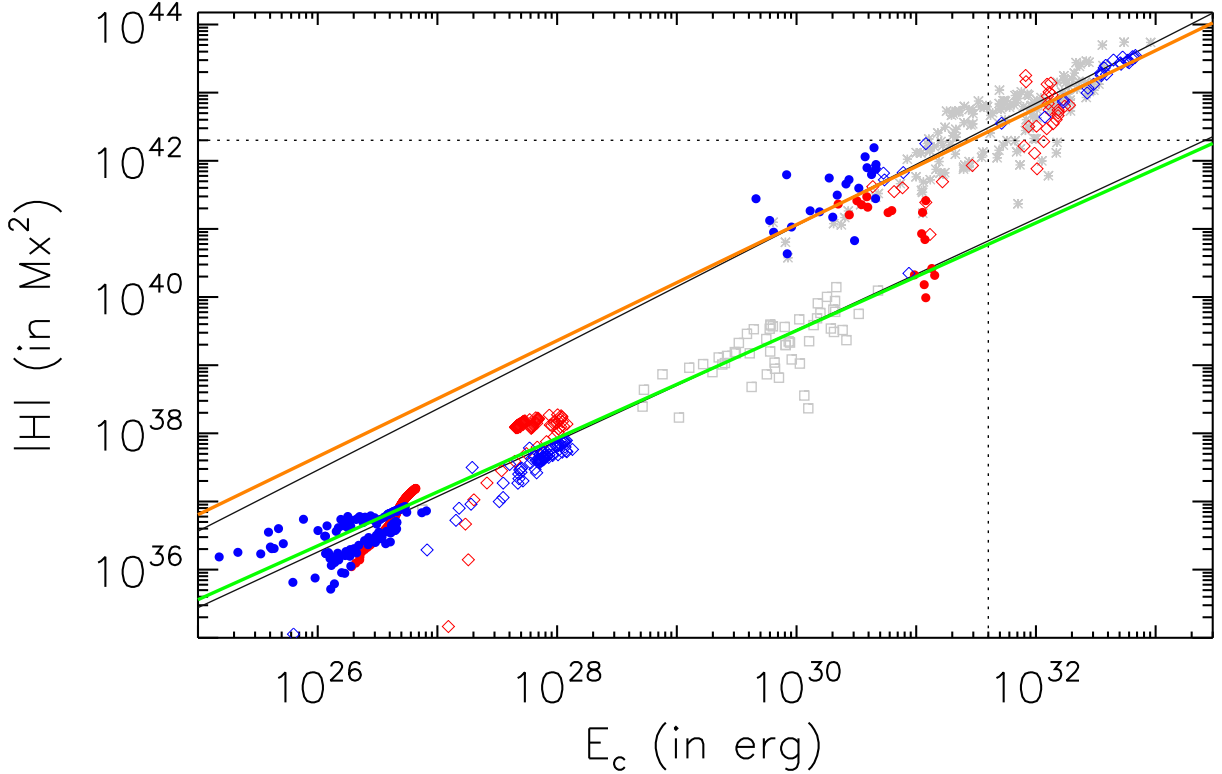}
\caption{Free magnetic energy -- relative helicity diagram for synthetic (lower left) and observed (upper right) ARs by means of their volume-calculated (red symbols) and surface-calculated (blue symbols) budgets. Eruptive (non-eruptive) AR cases are denoted by diamonds (circles).
The respective least-squares best fits are denoted by the thick orange (observed ARs) and green (MHD-simulated ARs) lines. Gray
symbols denote previous results on the EH diagram with asterisks corresponding to ARs \citep{tgr12} and squares
corresponding to quiet-Sun regions \citep{tz14}; their respective least-squares best fits are denoted by solid black lines.
The dotted horizontal and vertical lines correspond to the previously reported \citep{tgr12} thresholds of free magnetic
energy ($4\times10^{31}$~erg) and relative magnetic helicity ($2\times10^{42}$~Mx$^2$) for ARs to enter eruptive territory.}\label{figenh}
\end{figure*}

\section{Data description}
\label{S-data}

For our analysis we use the same data sets as in \cite{mor14}. These consist of timeseries of a) two synthetic MHD cases, a non-eruptive and an eruptive one, derived with three-dimensional, time-dependent, and resistive numerical MHD experiments \cite[e.g.,][]{arch14}, and b) two observed ARs, the non-eruptive NOAA AR 11072, and the eruptive NOAA AR 11158. Photospheric vector magnetic field observations of the latter were acquired with the Helioseismic and Magnetic Imager \citep[HMI;][]{sche12} onboard the Solar Dynamics Observatory \citep[SDO;][]{pes12}, while their three dimensional magnetic field configuration was derived with the nonlinear force-free field extrapolation method of \cite{wieg04}. The azimuthal 180\degr\ ambiguity in the photospheric SDO/HMI vector magnetograms was resolved by the nonpotential field
calculation (NPFC) method \citep{geo05,metc06}.

To derive the instantaneous free magnetic energy and relative magnetic helicity budgets with the first surface-calculation NLFF method (hereafter surface method), we use the synthetic lower-boundary and observed photospheric vector magnetograms for the MHD and observed cases respectively. On the other hand, to derive the energy/helicity budgets with the classical method (hereafter volume method), we use the corresponding three-dimensional field for the MHD cases and the extrapolated three-dimensional field for the observed ARs. In the present analysis, we do not comment on the temporal evolution of energy and helicity in models and observed data, but we treat each calculated budget as part of an independent pair of energy and helicity values for use in the EH diagram.

\section{Energy-helicity diagram of synthetic and observed cases}
\label{S-enhel}

Similarly to \cite{tgr12}, and \cite{tgl13}, we now construct the EH diagram for the two pairs of observed and synthetic ARs. Both the semi-analytical, volume-calculated and the NLFF results are included in the plot. The results are shown in Figure~\ref{figenh}.

\cite{tgr12} worked on a data set of 42 different observed ARs and reported a scaling relation $|H| \propto E_\mathrm{c}^{0.897}$ between the amplitude $|H|$ of the total relative helicity and the free energy $E_\mathrm{c}$ in the EH diagram.
For quiet-Sun regions, the respective scaling reported by \cite{tz14} was $|H| \propto E_\mathrm{c}^{0.815}$.
For the two observed ARs of this study, and including both NLFF and volume-calculation methods, we find
\begin{equation}
|H| \approx 3.319 \times 10^{15} E^{0.851}_{\rm c}\;\;,
\label{bestobs}
\end{equation}
while the best fit for the synthetic ARs gives
\begin{equation}
|H| \approx 6.45 \times 10^{15} E^{0.79}_{\rm c}.
\label{bestmhd}
\end{equation}
Evidently, therefore, both NLFF and volume-calculation methods for both observed and synthetic data give rise to a very similar EH scaling between the relative magnetic helicity and the free magnetic energy.
The global nature of this free-energy/relative-helicity scaling is reported here for the first time and warrants additional
investigation. As already stressed by \cite{tgr12}, however, scatter in the EH diagram should not be attributed solely to
numerical effects and uncertainties: deviations are expected in case of confined eruptions in ARs, where magnetic energy is
released, but relative helicity is roughly conserved. The extent and impact of this effect on the above EH scaling may be the subject of a future investigation.

In other findings from Figure~\ref{figenh}, notice that (i) the eruptive NOAA AR 11158 shows relative helicity and free energy above the eruptive thresholds reported by \cite{tgr12}, while the respective budgets for the noneruptive NOAA AR 11072 fall below these thresholds, and (ii) a similar segregation between the eruptive and the noneruptive cases exists for the synthetic data, as well.

\section{Discussion and conclusions}
\label{S-discuss}

We have constructed a comprehensive EH diagram of solar ARs using a) two different NLFF methods, namely a classical volume-calculation method (Eqs.~\ref{ecdef} and \ref{hmdef}) that derives the free magnetic energy and relative helicity budgets from the three-dimensional magnetic field, and a surface-calculation method (Eqs.~\ref{Ec_fin} and \ref{Hm_fin}) that is essentially a magnetic connectivity-based calculation of these two budgets from a single photospheric vector magnetogram, and b) both synthetic three-dimensional MHD and observed/extrapolated ARs. As already mentioned, both methods and both data sets show (i) a very similar scaling between relative magnetic helicity and free magnetic energy, such as the one derived by \cite{tgr12} for ARs, \cite{tgl13} for NOAA AR 11158, and \cite{tz14} for quiet-Sun regions, with a scaling index 0.84$\pm$0.05,
(ii) the existence of thresholds in both free energy and relative helicity for eruptive behavior in observed ARs \citep{tgr12,tgl13}, and (iii) a similar segregation  between eruptive and non-eruptive helicity/energy budgets for synthetic MHD data. It should be noted that the scaling between free magnetic energy and relative helicity holds despite the difference by a factor of up to $\sim3$ \citep{mor14} between respective budgets derived by the two methods. These findings  further attest to the robustness of the scaling relation between free magnetic energy and relative magnetic helicity in solar magnetic structures and provide a strong validity assessment of the surface NLFF method as compared to the classical energy and helicity calculation method.

There is significant scatter in the EH diagram, especially as free energy and relative-helicity values become lower (i.e., for the noneruptive MHD and observed cases). Apart from the occurrence of confined magnetic-reconnection events, where energy is dissipated but relative helicity is roughly conserved, resulting in departures from the scaling law, the scatter could also be attributed to a) uncertainties in the surface and volume calculations, b) the marginal lower boundary response to dynamic changes of the three-dimensional coronal field as a result of AR evolution or eruptions, and c) the incoherent nature of helicity for the weaker flux, noneruptive AR, and quiet-Sun cases. This incoherence, in terms of relative
helicity, was briefly discussed by \cite{tz14} for quiet-Sun magnetic structures, can be seen in Fig.~\ref{figenh} as a step change
in relative helicity as opposed to a smoother progression toward lower values in the free energy, compared to ARs. This helicity
incoherence can be attributed to the lack of a dominant helicity sense in quiet-Sun regions and will be investigated in more detail in a forthcoming work.

The lower value part of the EH diagram, corresponding to synthetic MHD cases, seems to be a rather continuous, smooth
progression in both free energy and relative helicity with respect to quiet-Sun values. We believe that this effect is incidental and should not be attributed to an incoherence of helicity in the MHD cases, but rather to the lower mean amplitude of magnetic field at the chosen lower boundary (middle of the photosphere) in these synthetic AR cases. These simulations - albeit corresponding to magnetized
plasma ($\beta$<1) even at photospheric heights - show, at this lower boundary, nearly two orders-of-magnitude weaker photospheric fields compared to observed ARs. Translated to magnetic helicity, weaker flux by a factor of 10$^2$ should correspond to smaller helicity amplitudes by a
factor of 10$^4$, hence the result of Fig.~\ref{figenh}. Nonetheless, the dominant sense of helicity in MHD cases is guaranteed by the physical
setup of the simulation, namely the buoyant emergence of a twisted flux tube.

In conclusion, we have reported in previous works \citep{tgr12, tgl13,tz14} that a robust scaling relation exists between the
free magnetic energy and the relative magnetic helicity in nonpotential solar magnetic structures. This scaling, reflected in the
relation |$H$| $\propto$ $E_{\rm c}^{0.84\pm0.05}$, has been adequately validated in this work and should be taken as a fact in future investigations. The pertaining question corresponds to the physical reasoning of this scaling, however. Possible links to the pre-eruption evolution of eruptive ARs \citep{tgl13}, the possible existence of an upper bound on helicity with respect to the photospheric flux and morphology \citep{zhan08}, the tendency of the majority of ARs to be noneruptive, or the similarities and distinctions between active regions and the quiet Sun \citep[e.g.,][]{sol03,aka14} should all be considered thoroughly, and may pave the way to further understand the nature and puzzles of solar magnetism.

\begin{acknowledgements}
We thank the referee, M.~A. Berger, for his support and encouragement. This research has been carried out in the framework of research projects  hosted by the RCAAM of the Academy of Athens. The observations used are courtesy of NASA/SDO and the HMI science team. We thank X.~Sun and Y.~Liu for the provided magnetic field extrapolations of NOAA AR 11158. The simulations were performed on the STFC and SRIF funded UKMHD cluster at the University of St Andrews. VA acknowledges support by the Royal Society. This work was supported from the EU's Seventh Framework Program under grant agreement n$^o$ PIRG07-GA-2010-268245. It has been also cofinanced by the European Union (European Social Fund -- ESF) and Greek national funds through the Operational Program ``Education and Lifelong Learning'' of the National Strategic Reference Framework (NSRF) -- Research Funding Program: Thales. Investing in knowledge society through the European Social Fund.
\end{acknowledgements}

\end{document}